\documentclass[twocolumn]{aastex62}

\usepackage[utf8]{inputenc}
\DeclareUnicodeCharacter{2212}{-}% support older LaTeX versions
\usepackage{graphicx}
\graphicspath{{./}{./figures/}{./tex/figures/}}
\usepackage{amsmath}
\usepackage{afterpage}
\usepackage{enumitem}
\usepackage{xcolor}
\setenumerate{itemsep=0mm}%noitemsep}

\usepackage{lineno}
\definecolor{mypink}{cmyk}{0, 0.7808, 0.4429, 0.1412}
\definecolor{newblue}{cmyk}{1,0.7,0,0}

\usepackage{soul} 
\usepackage{amsmath}
\usepackage{amssymb}
\usepackage{xspace}
\usepackage{xifthen}
\usepackage{eso-pic}

\newcommand{\habanero}{{Centaurus~I}\xspace}
\newcommand{\cayenne}{{DELVE~1}\xspace}

\newcommand{\Gaia}{{\it Gaia}\xspace}

\definecolor{forestgreen}{HTML}{228B22}
\definecolor{urlblue}{HTML}{000000}

\mathchardef\mhyphen="2D

\newcommand{\roughly}{\ensuremath{ {\sim}\,} }

\newlength{\dhatheight}

\newcommand{\code}[1]{\texttt{#1}\xspace}

\newcommand{\unit}[1]{\ensuremath{\mathrm{\,#1}}\xspace}
\newcommand{\yr}{\unit{yr}}
\newcommand{\Gyr}{\unit{Gyr}}

\newcommand{\degree}{\ensuremath{{}^{\circ}}\xspace}

\newcommand{\mas}{\unit{mas}}
\newcommand{\amin}{\unit{arcmin}}

\newcommand{\km}{\unit{km}}
\newcommand{\kms}{\km \second^{-1}}
\newcommand{\pc}{\unit{pc}}
\newcommand{\kpc}{\unit{kpc}}
\newcommand{\second}{\unit{s}}

\newcommand{\Msun}{\unit{M_\odot}}

\newcommand{\magn}{\unit{mag}}
\newcommand{\mmag}{\unit{mmag}}

\newcommand{\secref}[1]{Section~\ref{sec:#1}}

\newcommand{\tabref}[1]{Table~\ref{tab:#1}}

\newcommand{\figref}[1]{Figure~\ref{fig:#1}}
\newcommand{\figrefs}[2]{Figures~\ref{fig:#1} and \ref{fig:#2}}

\newcommand{\bandvar}[2][]{%
  \ifthenelse{\isempty{#1}}{\var{#2}}{\var{#2\_#1}}%
}

\newcommand{\modulus}{\ensuremath{m - M}\xspace}

\newcommand{\ra}{{\ensuremath{\alpha_{2000}}}\xspace}
\newcommand{\dec}{{\ensuremath{\delta_{2000}}}\xspace}
\newcommand{\age}{{\ensuremath{\tau}}\xspace}
\newcommand{\metal}{{\ensuremath{Z}}\xspace}

\newcommand{\feh}{{\ensuremath{\rm [Fe/H]}}\xspace}
\newcommand{\ellip}{\ensuremath{\epsilon}\xspace}
\newcommand{\PA}{\ensuremath{\mathrm{P.A.}}\xspace}
\newcommand{\TS}{\ensuremath{\mathrm{TS}}\xspace}

\newcommand{\HEALPix}{\code{HEALPix}}
\newcommand{\healpix}{\HEALPix}

\newcommand{\emcee}{\code{emcee}}
\newcommand{\ugali}{\code{ugali}}
\newcommand{\simple}{\code{simple}}
\newcommand{\var}[1]{\ensuremath{\texttt{\MakeUppercase{#1}}}\xspace}
\newcommand{\nside}{\code{nside}}

\providecommand\physrep{\ref@jnl{Phys.~Rep.}}%
\providecommand\apjs{\ref@jnl{ApJS}}%
\providecommand{\jcap}{\ref@jnl{JCAP}}%
\shorttitle{Two Ultra-Faint Milky Way Stellar Systems Discovered in DELVE}
\shortauthors{Mau \& Cerny et al.}

\begin{document}

\reportnum{\footnotesize FERMILAB-PUB-19-584-AE}

\title{Two Ultra-Faint Milky Way Stellar Systems Discovered in Early Data from the DECam Local Volume Exploration Survey}

% Author list file generated with: mkauthlist 1.2.3 
% mkauthlist authors.csv 

\author[0000-0003-3519-4004]{S.~Mau}
\affiliation{Kavli Institute for Cosmological Physics, University of Chicago, Chicago, IL 60637, USA}
\affiliation{Department of Astronomy and Astrophysics, University of Chicago, Chicago IL 60637, USA}
\author[0000-0003-1697-7062]{W.~Cerny}
\affiliation{Kavli Institute for Cosmological Physics, University of Chicago, Chicago, IL 60637, USA}
\affiliation{Department of Astronomy and Astrophysics, University of Chicago, Chicago IL 60637, USA}
\author[0000-0002-6021-8760]{A.~B.~Pace}
\affiliation{McWilliams Center for Cosmology, Carnegie Mellon University, 5000 Forbes Avenue, Pittsburgh, PA 15213, USA}
\author[0000-0003-1680-1884]{Y.~Choi}
\affiliation{Space Telescope Science Institute, 3700 San Martin Drive, Baltimore, MD 21218, USA}
\author[0000-0001-8251-933X]{A.~Drlica-Wagner}
\affiliation{Fermi National Accelerator Laboratory, P.O.\ Box 500, Batavia, IL 60510, USA}
\affiliation{Kavli Institute for Cosmological Physics, University of Chicago, Chicago, IL 60637, USA}
\affiliation{Department of Astronomy and Astrophysics, University of Chicago, Chicago IL 60637, USA}
\author[0000-0003-3402-6164]{L.~Santana-Silva}
\affiliation{Federal University of Rio de Janeiro, Valongo Observatory, Ladeira Pedro Antonio, 43, Sa\'ude 20080-090 Rio de Janeiro, Brazil}
\author[0000-0001-5805-5766]{A.~H.~Riley}
\affiliation{George P. and Cynthia Woods Mitchell Institute for Fundamental Physics and Astronomy, and Department of Physics and Astronomy, Texas A\&M University, College Station, TX 77843, USA}
\author[0000-0002-8448-5505]{D.~Erkal}
\affiliation{Department of Physics, University of Surrey, Guildford GU2 7XH, UK}
\author[0000-0003-1479-3059]{G.~S.~Stringfellow}
\affiliation{Center for Astrophysics and Space Astronomy, University of Colorado, 389 UCB, Boulder, CO 80309-0389, USA}
\author[0000-0002-6904-359X]{M.~Adam\'ow}
\affiliation{National Center for Supercomputing Applications, University of Illinois, 1205 West Clark Street, Urbana, IL 61801, USA}
\author[0000-0002-3936-9628]{J.~L.~Carlin}
\affiliation{LSST, 950 North Cherry Avenue, Tucson, AZ, 85719, USA}
\author[0000-0002-4588-6517]{R.~A.~Gruendl}
\affiliation{Department of Astronomy, University of Illinois, 1002 W. Green Street, Urbana, IL 61801, USA}
\affiliation{National Center for Supercomputing Applications, University of Illinois, 1205 West Clark Street, Urbana, IL 61801, USA}
\author{D.~Hernandez-Lang}
\affiliation{University of La Serena, La Serena, Chile}
\affiliation{Cerro Tololo Inter-American Observatory, NSF's National Optical-Infrared Astronomy Research Laboratory, Casilla 603, La Serena, Chile}
\affiliation{Gemini Observatory, La Serena, Chile}
\author[0000-0003-2511-0946]{N.~Kuropatkin}
\affiliation{Fermi National Accelerator Laboratory, P.O.\ Box 500, Batavia, IL 60510, USA}
\author[0000-0002-9110-6163]{T.~S.~Li}
\altaffiliation{NHFP Einstein Fellow}
\affiliation{Observatories of the Carnegie Institution for Science, 813 Santa Barbara Street, Pasadena, CA 91101, USA}
\affiliation{Department of Astrophysical Sciences, Princeton University, Princeton, NJ 08544, USA}
\author[0000-0002-9144-7726]{C.~E.~Mart\'inez-V\'azquez}
\affiliation{Cerro Tololo Inter-American Observatory, NSF's National Optical-Infrared Astronomy Research Laboratory, Casilla 603, La Serena, Chile}
\author{E.~Morganson}
\affiliation{National Center for Supercomputing Applications, University of Illinois, 1205 West Clark Street, Urbana, IL 61801, USA}
\author[0000-0001-9649-4815]{B.~Mutlu-Pakdil}
\affiliation{Department of Astronomy/Steward Observatory, 933 North Cherry Avenue, Room N204, Tucson, AZ 85721-0065, USA}
\author[0000-0002-7357-0317]{E.~H.~Neilsen}
\affiliation{Fermi National Accelerator Laboratory, P.O.\ Box 500, Batavia, IL 60510, USA}
\author[0000-0002-1793-3689]{D.~L.~Nidever}
\affiliation{Department of Physics, Montana State University, P.O. Box 173840, Bozeman, MT 59717-3840, USA}
\affiliation{NSF's National Optical-Infrared Astronomy Research Laboratory, 950 N. Cherry Ave., Tucson, AZ 85719, USA}
\author[0000-0002-7134-8296]{K.~A.~G.~Olsen}
\affiliation{NSF's National Optical-Infrared Astronomy Research Laboratory, 950 N. Cherry Ave., Tucson, AZ 85719, USA}
\author[0000-0003-4102-380X]{D.~J.~Sand}
\affiliation{Department of Astronomy/Steward Observatory, 933 North Cherry Avenue, Room N204, Tucson, AZ 85721-0065, USA}
\author[0000-0002-9599-310X]{E.~J.~Tollerud}
\affiliation{Space Telescope Science Institute, 3700 San Martin Drive, Baltimore, MD 21218, USA}
\author{D.~L.~Tucker}
\affiliation{Fermi National Accelerator Laboratory, P.O.\ Box 500, Batavia, IL 60510, USA}
\author[0000-0002-9541-2678]{B.~Yanny}
\affiliation{Fermi National Accelerator Laboratory, P.O.\ Box 500, Batavia, IL 60510, USA}
\author{A.~Zenteno}
\affiliation{Cerro Tololo Inter-American Observatory, NSF's National Optical-Infrared Astronomy Research Laboratory, Casilla 603, La Serena, Chile}
\author[0000-0002-7069-7857]{S.~Allam}
\affiliation{Fermi National Accelerator Laboratory, P.O.\ Box 500, Batavia, IL 60510, USA}
\author[0000-0001-5547-3938]{W.~A.~Barkhouse}
\affiliation{Department of Physics and Astrophysics, University of North Dakota, Grand Forks, ND 58202, USA}
\author[0000-0001-8156-0429]{K.~Bechtol}
\affiliation{Department of Physics, University of Wisconsin-Madison, Madison, WI 53706, USA}
\author[0000-0002-5564-9873]{E.~F.~Bell}
\affiliation{Department of Astronomy, University of Michigan, 1085 S.\ University Avenue, Ann Arbor, MI 48109, USA}
\author[0000-0002-8462-048X]{P.~Balaji}
\affiliation{Kavli Institute for Cosmological Physics, University of Chicago, Chicago, IL 60637, USA}
\affiliation{Department of Astronomy and Astrophysics, University of Chicago, Chicago IL 60637, USA}
\author[0000-0002-1763-4128]{D.~Crnojevi\'c}
\affiliation{Department of Chemistry and Physics, University of Tampa, 401 West Kennedy Boulevard, Tampa, FL 33606, USA}
\affiliation{Department of Physics \& Astronomy, Texas Tech University, Box 41051, Lubbock, TX 79409, USA}
\author{J.~Esteves}
\affiliation{Department of Physics, Brandeis University, Waltham, MA 02453, USA}
\author[0000-0001-6957-1627]{P.~S.~Ferguson}
\affiliation{George P. and Cynthia Woods Mitchell Institute for Fundamental Physics and Astronomy, and Department of Physics and Astronomy, Texas A\&M University, College Station, TX 77843, USA}
\author{C.~Gallart}
\affiliation{Instituto de Astrof\'{i}sica de Canarias, La Laguna, Tenerife, Spain}
\affiliation{Departamento de Astrof\'{i}sica, Universidad de La Laguna, Tenerife, Spain}
\author[0000-0002-1718-0402]{A.~K.~Hughes}
\affiliation{Department of Astronomy/Steward Observatory, 933 North Cherry Avenue, Room N204, Tucson, AZ 85721-0065, USA}
\author[0000-0001-5160-4486]{D.~J.~James}
\affiliation{Center for Astrophysics, Harvard \& Smithsonian, 60 Garden Street, Cambridge, MA 02138, USA}
\affiliation{Black Hole Initiative at Harvard University, 20 Garden Street, Cambridge, MA 02138, USA}
\author[0000-0003-0010-8129]{P.~Jethwa}
\affiliation{Institute for Astrophysics, T{\"u}rkenschanzstra{\ss}e 17, A-1180 Wien, Austria}
\author[0000-0001-6421-0953]{L.~C.~Johnson}
\affiliation{Center for Interdisciplinary Exploration and Research in Astrophysics (CIERA) and Department of Physics and Astronomy, Northwestern University, 2145 Sheridan Road, Evanston, IL 60208 USA}
\author[0000-0003-0120-0808]{K.~Kuehn}
\affiliation{Lowell Observatory, 1400 W Mars Hill Road, Flagstaff, AZ 86001, USA}
\affiliation{Australian Astronomical Optics, Macquarie University, North Ryde, NSW 2113, Australia}
\author[0000-0003-2025-3147]{S.~Majewski}
\affiliation{Department of Astronomy, University of Virginia, Charlottesville, VA, 22904, USA}
\author[0000-0002-1200-0820]{Y.-Y.~Mao}
\altaffiliation{NHFP Einstein Fellow}
\affiliation{Department of Physics and Astronomy, Rutgers, The State University of New Jersey, Piscataway, NJ 08854, USA}
\author[0000-0002-8093-7471]{P.~Massana}
\affiliation{Department of Physics, University of Surrey, Guildford GU2 7XH, UK}
\author[0000-0001-5435-7820]{M.~McNanna}
\affiliation{Department of Physics, University of Wisconsin-Madison, Madison, WI 53706, USA}
\author[0000-0003-2325-9616]{A.~Monachesi}
\affiliation{Instituto de Investigaci\'on Multidisciplinar en Ciencia y Tecnolog\'ia, Universidad de La Serena, Ra\'ul Bitr\'an 1305, La Serena, Chile}
\affiliation{Departmento de Astronom\'ia, Universidad de La Serena, Av. Juan Cisternas 1200 Norte, La Serena, Chile}
\author[0000-0002-1182-3825]{E.~O.~Nadler}
\affiliation{Department of Physics, Stanford University, 382 Via Pueblo Mall, Stanford, CA 94305, USA}
\affiliation{Kavli Institute for Particle Astrophysics \& Cosmology, P.O.\ Box 2450, Stanford University, Stanford, CA 94305, USA}
\author[0000-0002-8282-469X]{N.~E.~D.~No\"el}
\affiliation{Department of Physics, University of Surrey, Guildford GU2 7XH, UK}
\author[0000-0002-6011-0530]{A.~Palmese}
\affiliation{Fermi National Accelerator Laboratory, P.O.\ Box 500, Batavia, IL 60510, USA}
\author[0000-0003-1339-2683]{F.~Paz-Chinchon}
\affiliation{National Center for Supercomputing Applications, University of Illinois, 1205 West Clark Street, Urbana, IL 61801, USA}
\author[0000-0001-9186-6042]{A.~Pieres}
\affiliation{Instituto de F\'\i sica, UFRGS, Caixa Postal 15051, Porto Alegre, RS - 91501-970, Brazil}
\affiliation{Laborat\'orio Interinstitucional de e-Astronomia - LIneA, Rua Gal. Jos\'e Cristino 77, Rio de Janeiro, RJ - 20921-400, Brazil}
\author{J.~Sanchez}
\affiliation{Fermi National Accelerator Laboratory, P.O.\ Box 500, Batavia, IL 60510, USA}
\author[0000-0003-2497-091X]{N.~Shipp}
\affiliation{Kavli Institute for Cosmological Physics, University of Chicago, Chicago, IL 60637, USA}
\affiliation{Department of Astronomy and Astrophysics, University of Chicago, Chicago IL 60637, USA}
\affiliation{Fermi National Accelerator Laboratory, P.O.\ Box 500, Batavia, IL 60510, USA}
\author{J.~D.~Simon}
\affiliation{Observatories of the Carnegie Institution for Science, 813 Santa Barbara Street, Pasadena, CA 91101, USA}
\author[0000-0001-6082-8529]{M.~Soares-Santos}
\affiliation{Department of Physics, Brandeis University, Waltham, MA 02453, USA}
\affiliation{Fermi National Accelerator Laboratory, P.O.\ Box 500, Batavia, IL 60510, USA}
\author[0000-0001-6584-6144]{K.~Tavangar}
\affiliation{Kavli Institute for Cosmological Physics, University of Chicago, Chicago, IL 60637, USA}
\affiliation{Department of Astronomy and Astrophysics, University of Chicago, Chicago IL 60637, USA}
\author[0000-0001-7827-7825]{R.~P.~van~der~Marel}
\affiliation{Space Telescope Science Institute, 3700 San Martin Drive, Baltimore, MD 21218, USA}
\affiliation{Center for Astrophysical Sciences, Department of Physics \& Astronomy, Johns Hopkins University, Baltimore, MD 21218, USA}
\author[0000-0003-4341-6172]{A.~K.~Vivas}
\affiliation{Cerro Tololo Inter-American Observatory, NSF's National Optical-Infrared Astronomy Research Laboratory, Casilla 603, La Serena, Chile}
\author[0000-0002-7123-8943]{A.~R.~Walker}
\affiliation{Cerro Tololo Inter-American Observatory, NSF's National Optical-Infrared Astronomy Research Laboratory, Casilla 603, La Serena, Chile}
\author[0000-0003-2229-011X]{R.~H.~Wechsler}
\affiliation{Department of Physics, Stanford University, 382 Via Pueblo Mall, Stanford, CA 94305, USA}
\affiliation{Kavli Institute for Particle Astrophysics \& Cosmology, P.O.\ Box 2450, Stanford University, Stanford, CA 94305, USA}
\affiliation{SLAC National Accelerator Laboratory, Menlo Park, CA 94025, USA}

\collaboration{(DELVE Collaboration)}

\correspondingauthor{Sidney Mau, William Cerny, Alex Drlica-Wagner}
\email{sidneymau@uchicago.edu, williamcerny@uchicago.edu, kadrlica@fnal.gov}

\begin{abstract}
We report the discovery of two ultra-faint stellar systems found in early data from the DECam Local Volume Exploration survey (DELVE).
The first system, \habanero (DELVE~J1238$-$\allowbreak4054), is identified as a resolved overdensity of old and metal-poor stars with a heliocentric distance of $D_{\odot} = 116.3_{-0.6}^{+0.6}\kpc$, a half-light radius of $r_h = 2.3_{-0.3}^{+0.4}\amin$, an age of $\tau > 12.85\Gyr$, a metallicity of $\metal = 0.0002_{-0.0002}^{+0.0001}$, and an absolute magnitude of $M_V = -5.55_{-0.11}^{+0.11}\magn$.
This characterization is consistent with the population of ultra-faint satellites and confirmation of this system would make \habanero one of the brightest recently discovered ultra-faint dwarf galaxies.
\habanero is detected in \Gaia DR2 with a clear and distinct proper motion signal, confirming that it is a real association of stars distinct from the Milky Way foreground; this is further supported by the clustering of blue horizontal branch stars near the centroid of the system.
The second system, \cayenne (DELVE~J1630$-$\allowbreak0058), is identified as a resolved overdensity of stars with a heliocentric distance of $D_{\odot} = 19.0_{-0.6}^{+0.5}\kpc$, a half-light radius of $r_h = 0.97_{-0.17}^{+0.24}\amin$, an age of $\tau = 12.5_{-0.7}^{+1.0}\Gyr$, a metallicity of $\metal = 0.0005_{-0.0001}^{+0.0002}$, and an absolute magnitude of $M_V = -0.2_{-0.6}^{+0.8}\magn$, consistent with the known population of faint halo star clusters.
Given the low number of probable member stars at magnitudes accessible with \Gaia DR2, a proper motion signal for \cayenne is only marginally detected.
We compare the spatial position and proper motion of both \habanero and \cayenne with simulations of the accreted satellite population of the Large Magellanic Cloud (LMC) and find that neither is likely to be associated with the LMC.
\end{abstract}


%-------------------------------------------------------------------------------

\section{Introduction}
\label{sec:intro}

Ultra-faint dwarf galaxies are the least luminous and most dark-matter-dominated objects in the known universe.
They are generally characterized by their low luminosities, relatively large mass-to-light ratios, and old, metal-poor stellar populations \citep[e.g.,][and references therein]{McConnachie:2012,PaperI,Simon:2019}.
%The most recently discovered ultra-faint galaxies around the Milky Way have been detected as arcminute-scale stellar overdensities whose color--magnitude diagrams are consistent with old, metal-poor isochrones in optical--near-infrared sky surveys using automated search algorithms 
As dark-matter-dominated systems, ultra-faint galaxies are among the most pristine laboratories for the study of dark matter itself.
For instance, they serve as excellent candidates for the indirect detection of dark matter annihilation and decay \citep[e.g.,][]{Geringer-Sameth:2015b,Albert:2017}.
Furthermore, the census of Milky Way satellite galaxies has been used to constrain models of particle dark matter (e.g., cold, warm, and self-interacting dark matter), which predict different structures at small scales \citep[e.g.,][]{Aaronson:1983,Maccio:2010,Lovell:2014,Jethwa:2018,Nadler:2019b}.
The demographics of the Milky Way satellite population have been used to test our understanding of reionization \citep[e.g.,][]{Boylan-Kolchin:2015}, the formation of the smallest galaxies \citep[e.g.,][]{Jeon:2017,Wheeler:2019}, the galaxy--halo connection \citep[e.g,][]{Jethwa:2018,Kim:2018,Newton:2018,Nadler:2019a}, and the origin of the heavy elements \citep[e.g.,][]{Ji:2016,Frebel:2018}.
As such, there has been great interest in the discovery, confirmation, and characterization of new faint systems.

Faint halo star clusters form another population of stellar systems in orbit around the Milky Way.
While their surface brightnesses are comparable to those of the ultra-faint galaxies, they are generally characterized by having smaller physical sizes ($r_{1/2} \lesssim 20\pc$) and heliocentric distances ($D_{\odot} \gtrsim 15\kpc$) than dark-matter-dominated satellite galaxies.
These faint star clusters are proposed to have been accreted onto the Milky Way through the disruption of infalling satellite galaxies \citep[e.g.,][]{Gnedin:1997,Searle:1978,Koposov:2007,Forbes:2010,Leaman:2013,Massari:2017}.
As such, understanding the population of faint halo star clusters is an important aspect in understanding the assembly history of the Milky Way.
%Despite their different physical sizes, faint halo star clusters and faint dwarf galaxies tend to have comparable surface brightnesses.
While physical sizes can be used as a proxy to categorize objects as either ultra-faint galaxies or faint star clusters, the most definitive classification comes from the kinematic measurement of dark matter content via spectroscopic analysis.

%In this context, wide-area digital sky surveys in optical--near-infrared wavelengths have resulted in the discovery of many new dwarf galaxies.
Before the advent of the Sloan Digital Sky Survey (SDSS), there were only a dozen known Milky Way satellite galaxies.
The unprecedented depth of SDSS over most of the northern sky resulted in a doubling of the known population of satellite galaxies during the decade from 2005 to 2015 \citep[e.g.,][]{Willman2005AJ....129.2692W, Willman:2005,Belokurov2006ApJ...647L.111B,Belokurov2007ApJ...654..897B, Belokurov2009MNRAS.397.1748B, 2010ApJ...712L.103B,ZuckerApJ...650L..41Z,Zucker2006ApJ...643L.103Z}.
By virtue of successive large sky surveys, including those using the Dark Energy Camera \citep[DECam;][]{Flaugher:2015} installed on the 4 m Blanco Telescope at the Cerro Tololo Inter-American Observatory (CTIO) in Chile, the current number of Milky Way satellite galaxies has increased to $\roughly 60$ in the past five years.
Simultaneously, new star clusters have been discovered at increasingly faint magnitudes, contributing to the overall population of stellar systems orbiting the Milky Way.
%\WC{these numbers refer to satelite galaxies specifically. need to specify 'galaxies'}
%\SM{This is a good point. It's difficult to classify some of these systems, so I think I've defaulted to ``satellites'' as a neutral solution, but the 60 number is more or less the census of dwarf galaxies.}
%\YC{I meant to talk about the number of MW satellite dwarf galaxies here. I put 'galaxies' to avoid any confusion.}
Specifically, searches for Milky Way satellites in the Dark Energy Survey \citep[DES; e.g.,][]{Bechtol:2015,Drlica-Wagner:2015,Kim:2015c,Koposov:2015cua,Luque:2016} and Pan-STARRS \citep[e.g.,][]{Laevens:2014,Laevens:2015a,Laevens:2015b} resulted in the discovery of more than 20 new satellites.
%\SM{Be careful here, some of these objects are now generally classified as star clusters, so quoting numbers seems a bit dangerous.}
Deep imaging surveys using the Hyper Suprime-Cam (HSC) have also uncovered three new candidate dwarf galaxies at distances and brightnesses inaccessible to previous surveys \citep{Homma:2016,Homma:2017,Homma:2019}.
Meanwhile, there have been a number of community-led DECam surveys that have contributed to the census of Milky Way satellites.
These include the Survey of the MAgellanic Stellar History \citep[SMASH; e.g.,][]{martin_2015_hydra_ii, Nidever2017AJ....154..199N}, the Magellanic SatelLites Survey \citep[MagLiteS; e.g.,][]{Drlica-Wagner:2016,Torrealba:2018a}, the Magellanic Edges Survey \citep[e.g.,][]{Koposov:2018}, and the Blanco Imaging of the Southern Sky Survey \citep[e.g.,][]{Mau:2019}.
With increasing sky coverage and depth, DECam is expected to continue to play an important role in searching for ultra-faint Milky Way satellites in the southern sky.
We refer the reader to \citet{Simon:2019} and references therein for a recent review of the Milky Way ultra-faint satellite galaxy population.

As a continuation of these community-led surveys in the southern hemisphere, the DECam Local Volume Exploration survey (DELVE)\footnote{\url{https://delve-survey.github.io}} seeks to complete DECam coverage of the southern sky with $|b| > 10$\degree by combining 126 nights of new observations in the 2019A--2021B semesters with existing public DECam community data.
DELVE consists of three survey components: a shallow wide-area survey of the southern sky (WIDE), a medium-depth survey around the Magellanic Clouds (MC; this serves as an extension of SMASH and MagLiteS), and a deep-drilling survey around four Magellanic analogs in the Local Volume (DEEP; e.g., similar to \citealt{Sand:2015} and \citealt{Carlin:2016}).
%In particular, DELVE-WIDE is designed to search for new ultra-faint stellar systems in the Milky Way, and it will entirely map the high-Galactic-latitude southern sky with a comparable depth to that of the first two years of DES after having completed its three years of observations.
In particular, DELVE-WIDE is designed to search for new ultra-faint stellar systems around the Milky Way by mapping the high-Galactic-latitude southern sky to a depth comparable to that of the first two years of DES.

Using an early version of the DELVE-WIDE catalog, which was constructed from existing public DECam exposures and DELVE exposures that were taken primarily in 2019A, we conducted a search for new faint Milky Way satellites and found two new resolved stellar overdensities that are consistent with old, metal-poor isochrones.
Furthermore, we cross-matched the early DELVE-WIDE data with the \Gaia DR2 catalog \citep{Gaia:2018} in the regions around these systems and measured their proper motions, helping confirm that these systems are real associations of stars.
The first of these systems, DELVE~J1238$-$4054, has a physical size and luminosity consistent with the locus of ultra-faint galaxies (\tabref{properties}, \figref{satellites}), and we tentatively denote it Centaurus~I (Cen~I).
In contrast, the small physical size and extremely low luminosity of the second system, DELVE~J1630$-$0058, are consistent with the population of faint halo star clusters (\tabref{properties}, \figref{satellites}), and we tentatively assign it the name DELVE~1.
We note that this system was simultaneously discovered in Pan-STARRS DR1 \citep{PaperI}.
While kinematic measurements are necessary to definitively classify the nature of faint stellar systems, this labeling scheme follows the convention of naming ultra-faint galaxies after the constellation in which they reside and faint star clusters after the survey in which they were discovered.
%We refer to these systems as DELVE~J1238$-$4054 and DELVE~J1630$-$0058 based on their positions in equatorial coordinates.
%Morphological and isochrone parameters fit to these systems suggest that DELVE~J1238$-$4054 is consistent with the population of ultra-faint dwarf galaxies, and DELVE~J1630$-$0058 is consistent with the population of ultra-faint outer halo star clusters.
%We tentatively attribute the names of Centaurus~I (Cen~I) and DELVE~1, respectively, to these systems, following the convention of naming UFD galaxies after the constellation in which they reside and faint star clusters after the survey in which they were discovered.

This paper is organized as follows.
In \secref{data}, we describe DELVE-WIDE and the early catalog used in this study.
The search algorithm is described in \secref{search}.
In \secref{results}, we present the morphology, isochrone parameters, and proper motions of \habanero and \cayenne.
Finally, in \secref{discussion}, we discuss interesting features of each system as well as their possible origins.
We briefly conclude in \secref{conclusion}.

%-------------------------------------------------------------------------------

\section{Data}
\label{sec:data}

DELVE-WIDE seeks to achieve complete, contiguous coverage of the high-Galactic-latitude ($|b| > 10\degree$) southern sky in $g,r,i,z$ by targeting regions of the sky that have not been observed by other community programs.
DELVE is expected to collect $\roughly 20{,}000$ new exposures over its three-year survey.
During the first year of DELVE observing, we performed $3 \times 90\second$ dithered exposures in $g$, $i$ following the survey strategy of DES.
We leave a detailed description of the DELVE observing and data reduction to a future paper.
%By the end of DELVE, we expect to achieve median $5 \sigma$ depths for point sources of $g \sim 23.4 \magn$ and $i \sim 22.4 \magn$.
%By combining devoted DELVE observations with archival DECam data, DELVE seeks to cover the southern hemisphere high-Galactic-latitude sky to a $5\sigma$ depth of $g = 24.0$, $r = 23.8$, $i = 23.1$, and $z = 22.8$.

%ABP: About the g/i strategy, you can mention that these bands were selected to maximum telescope time (dark versus bright/gray) and satellite galaxy identification.  g/i is only slightly worse than g/r.  However, since the region of sky we searched over in the first data release was targeted by e-rosita in r/z we used g/r.
%, reaching 10$\sigma$ depths of g $\sim$ 23.5 and i $\sim$ 22.7 (estimated from DES and MagLiteS)

Our early DELVE-WIDE dataset consists of approximately 14,000 exposures in the northern Galactic cap with $b > 10\degree$ and $\delta_{2000} < 0\degree$.
%This dataset includes DELVE observations in $gi$ ($\sim$6,200 exposures) and other DECam exposures in $griz$ ($\sim$7,800 exposures) that were publicly available in August 2019.\footnote{Public exposures were downloaded from the Science Archive hosted by NSF's National Optical-Infrared Astronomy Research Laboratory: \url{http://archive1.dm.noao.edu/}.} 
% \ADW{What fraction are new DELVE exposures?}\YC{If I remember correctly, 5k comes from DELVE gi. William can confirm this.}
The main constituents of this dataset are observations taken by DELVE, DECaLS \citep{Dey:2019}, and DeROSITAS,\footnote{\url{http://astro.userena.cl/derositas/}} augmented by other DECam exposures in $griz$ that were publicly available in 2019 August.\footnote{Public exposures were downloaded from the Science Archive hosted by NSF's National Optical-Infrared Astronomy Research Laboratory: \url{http://archive1.dm.noao.edu}.}
All exposures were processed consistently with the DES Data Management (DESDM) pipeline \citep{Morganson:2018}.
This pipeline enables sub-percent-level photometric accuracy by calibrating based on custom-made, seasonally averaged bias and flat images and performing full-exposure sky background subtraction \citep{Bernstein:2018}.
The DESDM pipeline utilizes \code{SourceExtractor} and \code{PSFEx} \citep{Bertin:1996,Bertin:2011} for automatic source detection and photometric measurement on an exposure-level basis.
We then calibrate stellar positions against \Gaia DR2 \citep{Gaia:2018}, which provides 30\mas astrometric calibration precision.
The DELVE photometry is calibrated by matching stars in each CCD to the APASS \citep{Henden:2014} and Two Micron All Sky Survey \citep[2MASS;][]{Skrutskie:2006} sky survey catalogs following the procedure described in \citet{Drlica-Wagner:2016}.
APASS-measured magnitudes were transformed to the DES filter system before calibration using the equations described in Appendix~A4 of \citet{Drlica-Wagner:2018}:
\small\begin{align*}
g_{\rm DES} &= g_{\rm APASS} - 0.0642(g_{\rm APASS}-r_{\rm APASS}) - 0.0239 \\
r_{\rm DES} &= r_{\rm APASS} - 0.1264(r_{\rm APASS}-i_{\rm APASS}) - 0.0098 \\
i_{\rm DES} &= r_{\rm APASS} - 0.4145(r_{\rm APASS}-J_{\rm 2MASS} - 0.81) - 0.0391,
\end{align*}\normalsize
which have statistical rms errors per star of $\sigma_g = 0.04\magn$, $\sigma_r = 0.05\magn$, and $\sigma_i = 0.04\magn$.
The relative photometric uncertainty of these derived zero points was estimated to be $\roughly 3\%$ by comparing to measurements made with the DES Forward Global Calibrations Module \citep[FGCM;][]{Burke:2018} in overlapping fields.
In a small number of cases where too few stars in a given exposure were matched with the reference catalog, we derived photometric zero points from a simultaneous fit of all CCDs for that exposure.

We built a multiband catalog of unique sources by matching detections between the individual single-exposure catalogs following \citet{Drlica-Wagner:2015}.
We started by selecting DECam exposures in the DELVE-WIDE dataset with exposure times ranging from 30 to 350\second.
We applied basic exposure-level cuts on the effective exposure time scale factor ($\var{t\_eff} > 0.3$; \citealt{Neilsen:2015}), astrometric matching quality vs. \Gaia ($\var{astrometric\_chi2} < 500$), and number of objects ($\var{n\_objects} < 7.5 \times 10^{5}$) to remove exposures that suffered from observational, instrumental, and/or processing artifacts.
To generate a unique source catalog with multiband information, we cross-matched all sources detected in individual exposures using a $1''$ matching radius.
We calculated weighted-average photometric properties based on the single-exposure measurements and their associated uncertainties \citep{Drlica-Wagner:2015}.
In total, the dataset covers approximately $6{,}000\deg^{2}$ in any single band, with the $g$ and $r$ bands providing the largest simultaneous coverage in any two bands.
There are 437,373,694 unique objects in this early catalog.

Extinction from Milky Way foreground dust was calculated for each object from a bilinear interpolation to the extinction maps of \citet{Schlegel:1998} and \citet{Schlafly:2011}.
To calculate reddening, we assumed $R_V = 3.1$ and used a set of $R_{\lambda} = A_{\lambda}/E(B-V)$ coefficients derived by DES for the $g$, $r$, and $i$ bands, where $R_g = 3.185$, $R_r = 2.140$, and $R_i = 1.571$, respectively \citep{DR1:2018}.\footnote{An update to the DECam standard bandpasses changed these coefficients by $< 1\mmag$ for DES DR1 \citep{DR1:2018}.}
Hereafter, all quoted magnitudes are corrected for dust extinction.

%-------------------------------------------------------------------------------

\section{Satellite Search}
\label{sec:search}

\begin{figure*}
\center
Diagnostic Plots for Centaurus~I\\
\includegraphics[width=\textwidth]{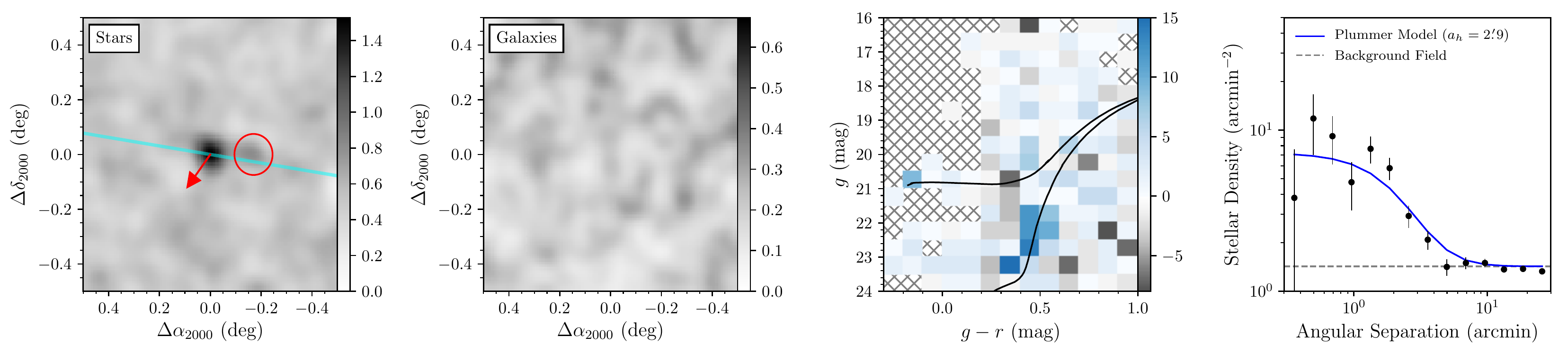}
\caption{
    Source density, color--magnitude diagram, and radial density profile plots for \habanero.
    (Left) Stellar density field convolved with a Gaussian kernel of 2\arcmin.
    The red arrow is drawn in the direction of the solar-reflex-corrected proper motion, and the cyan line corresponds to the great circle connecting \habanero and the Galactic center.
    A secondary overdensity near \habanero, which is a potential tidal feature, is circled in red.
    (Middle left) Background galaxy density field convolved with a Gaussian kernel of 2\arcmin.
    (Middle right) Color--magnitude Hess diagram corresponding to all foreground stars within 0\fdg10 of the centroid of \habanero minus all background stars in a concentric annulus from 0\fdg24 to 0\fdg26.
    The best-fit \code{PARSEC} isochrone (derived in \secref{morph}; \tabref{properties}) is shown in black.
    Crosshatching indicates bins with no stars.
    (Right) Radial surface density profile of stars passing the isochrone filter; the errors are derived from the standard deviation of the number of stars in a given annulus divided by the area of that annulus.
    The blue curve corresponds to the best-fit Plummer model, assuming spherical symmetry, with $a_h=2\farcm9$ (derived in \secref{morph}; \tabref{properties}).
    The dashed gray line represents the background field density.
}
\label{fig:habanero_diagnostic}

Diagnostic Plots for DELVE~1\\
\includegraphics[width=\textwidth]{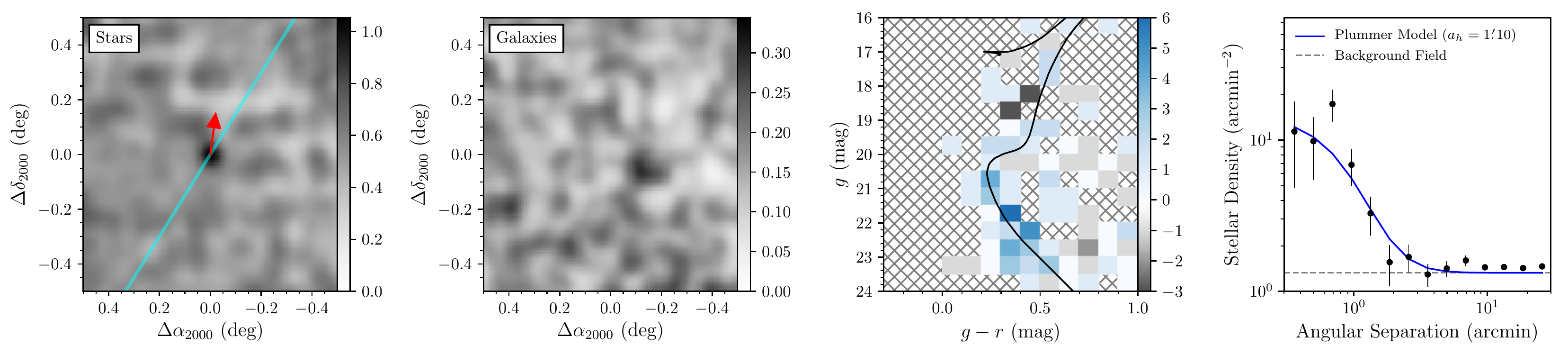}
\caption{
    Similar to \figref{habanero_diagnostic} but for \cayenne.
    Due to incomplete coverage in this region, CCD chip gaps (i.e., the underdense horizontal and vertical striations) are apparent in the left panel.
}
\label{fig:cayenne_diagnostic}
\end{figure*}

To identify Milky Way satellite candidates in the early DELVE-WIDE catalog, we applied the \simple\footnote{\url{https://github.com/DarkEnergySurvey/simple}} algorithm, which has successfully been used for satellite searches on other DECam and Pan-STARRS datasets \citep[e.g.,][]{Bechtol:2015,PaperI,Mau:2019}.
Briefly, \simple uses an isochrone filter in the color--magnitude space of two bands to enhance the contrast of halo substructures relative to the foreground field of Milky Way stars at a given small range of distances.
Because the total area covered in both the $r$ and $i$ bands is roughly equal for the DELVE-WIDE catalog, we chose to run \simple using $g$- and $r$-band data.
The $r$ band was chosen over the $i$ band because it was found to be deeper.
%has a $10 \sigma$ point-source depth that is \CHECK{$\roughly 0.8\magn$ deeper than that of the $i$-band in this catalog}.
%This follows \citet{Bechtol:2015,Drlica-Wagner:2015}.
We note that running \simple on $g$- and $i$-band data yields similar findings and also results in the detections of both systems presented in this paper at high significance.

Stars were selected with $|\var{spread\_model\_r}| < 0.003 + \var{spreaderr\_model\_r}$, where $\var{spread\_model}$ is a morphological variable acting as a discriminant between the best-fitting local point-spread function (PSF) model (for a point-like source) and the same PSF model but convolved with a circular exponential disk model with a scale length of one-sixteenth of the PSF's FWHM (for an extended source), and $\var{spreaderr\_model}$ is the associated error \citep{2012ApJ...757...83D,DR1:2018}.
A magnitude selection of $g < 23\magn$ was applied to reduce star--galaxy confusion.

%Spatial overdensities of old, metal-poor stars were identified with a matched-filter isochrone, scanning in distance modulus from 16.0\magn to 23.0\magn in steps of 0.5\magn.
%Specifically, a \code{PARSEC} isochrone \citep{Bressan:2012} with metallicity $\metal=0.0001$ and age $\age=12\Gyr$ was used.
%At each step in the distance modulus scan, stars were selected within 0.1\magn of the isochrone locus in color--magnitude space according to $\Delta (g-r) < \sqrt{0.1^2 + \sigma_g^2 + \sigma_r^2}$, where $\sigma_g$ and $\sigma_r$ are the photometric uncertainties on the $g$- and $r$-band magnitudes, respectively.
%
%The catalog was divided into \healpix \citep{Gorski:2005} pixels of $\nside=32$ ($\roughly 3.4\deg^2$).
%For each $\nside=32$ pixel and distance modulus step, the isochrone filter described above was applied to create a map of the filtered stellar density field, which was then smoothed by a Gaussian kernel ($\sigma = 2\arcmin$).
%Local density peaks were identified by iteratively raising a density threshold until fewer than ten disconnected peaks remained above the threshold value.
%For each identified peak, the Poisson significance of the observed stellar counts relative to the local field density within a given aperture was computed.
%All peaks with Poisson significance ${\rm SIG} > 5.5 \sigma$ were considered seeds for subsequent analysis.

The DELVE-WIDE catalog was divided into \healpix \citep{Gorski:2005} pixels of $\nside=32$ ($\roughly 3.4\deg^2$).
For each $\nside=32$ pixel, spatial overdensities of old, metal-poor stars were identified with a matched-filter isochrone, scanning in distance modulus from 16.0 to 23.0\magn in steps of 0.5\magn.
Specifically, a \code{PARSEC} isochrone \citep{Bressan:2012} with metallicity $\metal=0.0001$ and age $\age=12\Gyr$ was used.
At each step in the distance modulus scan, stars were selected within 0.1\magn of the isochrone locus in color--magnitude space according to $\Delta (g-r) < \sqrt{0.1^2 + \sigma_g^2 + \sigma_r^2}$, where $\sigma_g$ and $\sigma_r$ are the photometric uncertainties on the $g$- and $r$-band magnitudes, respectively.
The map of the filtered stellar density field was then smoothed by a Gaussian kernel ($\sigma = 2\arcmin$), and local density peaks were identified by iteratively raising a density threshold until fewer than 10 disconnected peaks remained above the threshold value.
For each identified peak, the Poisson significance of the observed stellar counts relative to the local field density within a given aperture was computed.
All peaks with Poisson significance ${\rm SIG} > 5.5 \sigma$ were considered for subsequent analysis.

Upon visual inspection of diagnostic plots for each of these peaks, two were identified as potential Milky Way satellite candidates, which we designate \habanero and \cayenne (\figrefs{habanero_diagnostic}{cayenne_diagnostic}, respectively).
%The left two panels of \figrefs{habanero_diagnostic}{cayenne_diagnostic} show the filtered and smoothed stellar and galactic density fields, respectively.
%The middle right panels show the color--magnitude Hess diagram of stars within 0\fdg10 of the centroid minus the stars in a concentric annulus of inner and outer radii of 0\fdg24 and 0\fdg26, respectively (i.e., the background-subtracted color--magnitude diagram); the best-fit \code{PARSEC} isochrone (derived in \secref{morph}; \tabref{properties}) is overplotted as a black line.
%The right panels show the radial distribution of isochrone-filtered stars with respect to the centroid; the best-fit Plummer profile (derived in \secref{morph}; \tabref{properties}) is shown in blue, and the background field estimate is shown as a dashed gray line.
%Note that \cayenne was discovered in a region of the survey with incomplete coverage, and CCD chip gaps can be seen in the upper right region of the left panel of \figref{cayenne_diagnostic}.
The left two panels of \figrefs{habanero_diagnostic}{cayenne_diagnostic} show the filtered and smoothed stellar and galactic density fields, respectively.
The middle right panels show the color--magnitude Hess diagram.
The right panels show the radial distribution of isochrone-filtered stars with respect to the centroid.
Note that \cayenne was discovered in a region of the survey with incomplete coverage, and CCD chip gaps can be seen in the upper-right region of the left panel of \figref{cayenne_diagnostic}.

%-------------------------------------------------------------------------------

\section{Properties of the Discovered Stellar Systems}
\label{sec:results}

In the following subsections, we characterize the morphologies, stellar populations, distances, and proper motions of \habanero and \cayenne.
The most probable values of these parameters, with associated uncertainties, are presented in \tabref{properties}.

\begin{figure*}
\center
Membership Plots for Centaurus~I\\
\includegraphics[width=\textwidth]{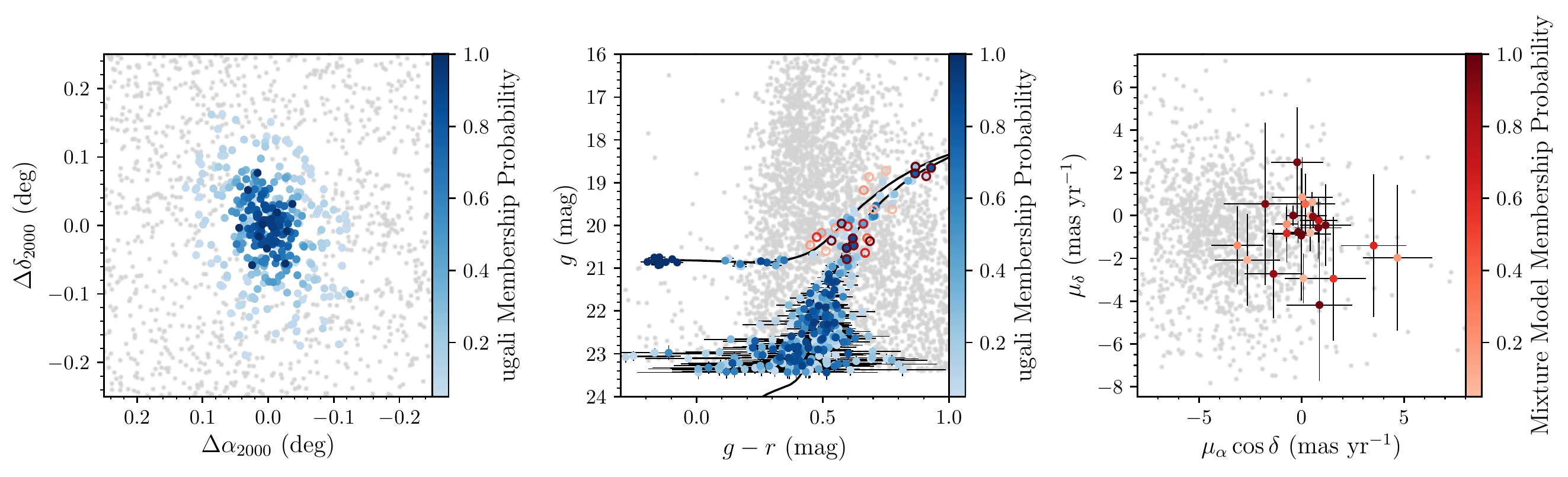}
\caption{
    Spatial distribution and color--magnitude diagram plots colored by \ugali membership probability ($p_{\rm \ugali}$) and proper motion plots colored by mixture model membership probability ($p_{\rm MM}$) for \habanero.
    (Left) Spatial distribution of stars with $g < 23.5\magn$ in a $0.25\deg^2$ area region around the centroid of \habanero.
    Stars with $p_{\rm \ugali} > 0.05$ are colored by their \ugali membership probability, and stars with $p_{\rm \ugali} \leq 0.05$ are shown in gray.
    (Center) Color--magnitude diagram of the stars shown in the left panel; the errors are derived from the photometric uncertainties of each band.
    The best-fit \code{PARSEC} isochrone (\tabref{properties}) is drawn in black.
    Several blue horizontal branch stars are identified as highly probable members of \habanero and are clustered very closely to the centroid of the system.
    Stars cross-matched with \Gaia DR2 with $p_{\rm MM} > 0.05$ are outlined by their mixture model membership probability.
    (Right) \Gaia proper motions for stars cross-matched with DELVE-WIDE.
    Stars with $p_{\rm MM} > 0.05$ are colored by their mixture model membership probability, and stars with $p_{\rm MM} \leq 0.05$ are shown in gray.
    The \Gaia signal for \habanero is distinct against the background field stars.
}
\label{fig:habanero_membership}

Membership Plots for DELVE~1\\
\includegraphics[width=\textwidth]{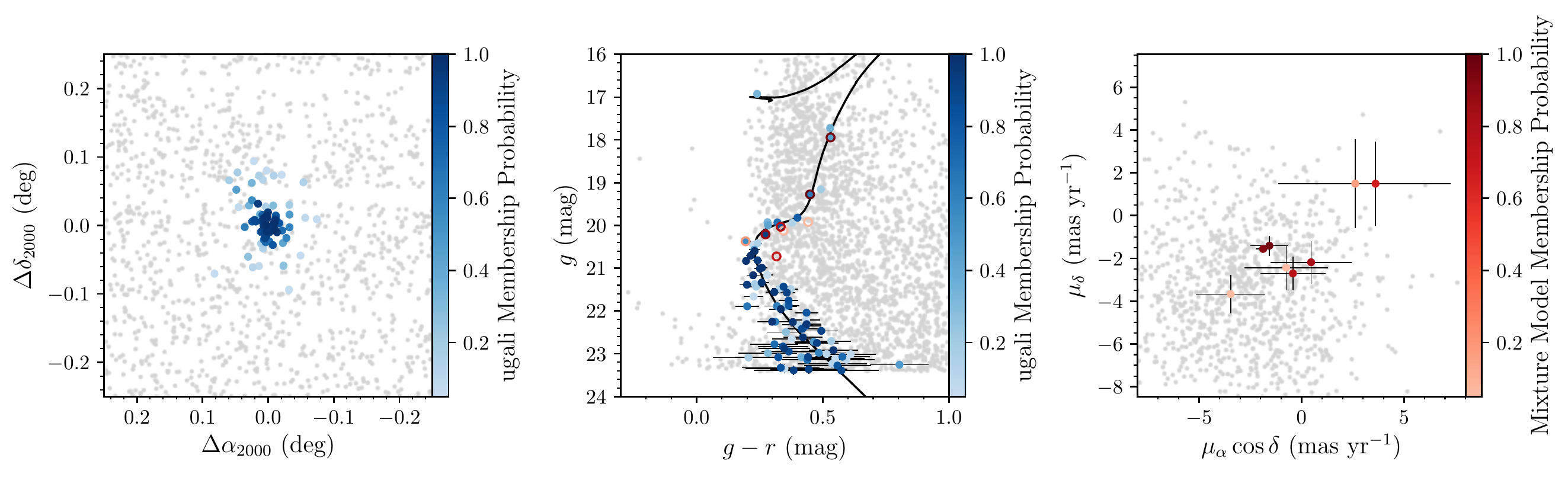}
\caption{
    Similar to \figref{habanero_membership} but for \cayenne.
    The \Gaia proper motion signal of \cayenne is marginally detected.
}
\label{fig:cayenne_membership}
\end{figure*}

\begin{deluxetable}{l c c}
\tablecolumns{3}
\tablewidth{0pt}
\tabletypesize{\footnotesize}
\tablecaption{\label{tab:properties}
Derived morphology, isochrone, and proper motion parameters for \habanero and \cayenne.
}
\tablehead{
\colhead{Parameter} & \colhead{Centaurus~I} & \colhead{DELVE~1}}
\startdata
\ra (deg) & $189.585^{+0.004}_{-0.004}$ & $247.725^{+0.002}_{-0.002}$ \\
\dec (deg) & $-40.902^{+0.004}_{-0.005}$ & $-0.972^{+0.003}_{-0.003}$ \\
$a_h$ (arcmin) & $2.9^{+0.5}_{-0.4}$ & $1.10^{+0.27}_{-0.19}$  \\
$r_h$ (arcmin) & $2.3^{+0.4}_{-0.3}$ & $0.97^{+0.24}_{-0.17}$  \\
$r_{1/2}$ (pc) & $79^{+14}_{-10}$ & $5.4^{+1.5}_{-1.1}$  \\
\ellip & $0.4^{+0.1}_{-0.1}$ & $0.2^{+0.1}_{-0.2}$  \\
\PA (deg) & $20^{+11}_{-11}$ & $21^{+26}_{-30}$  \\
\modulus (mag) & $20.33^{+0.03}_{-0.01} \pm 0.1$\tablenotemark{a} & $16.39^{+0.06}_{-0.07} \pm 0.1$\tablenotemark{a}  \\
$D_{\odot}$ (kpc) & $116.3^{+1.6}_{-0.6}$ & $19.0^{+0.5}_{-0.6}$ \\
\age (Gyr) & $>12.85$\tablenotemark{b} & $12.5^{+1.0}_{-0.7}$  \\
\metal & $0.0002^{+0.0001}_{-0.0002}$ & $0.0005^{+0.0002}_{-0.0001}$  \\
$\sum_i p_{i, {\rm \ugali}}$ & $155^{+19}_{-20}$ & $50^{+8}_{-9}$  \\
\TS & $308.3$ & $146.7$ \\[-0.5em]
\multicolumn{3}{c}{\hrulefill} \\
$M_V$ (mag) & $-5.55^{+0.11}_{-0.11}$\tablenotemark{c} & $-0.2^{+0.8}_{-0.6}$\tablenotemark{c}  \\
$M_{*}$ (${\rm M}_{\odot}$) & $14300^{+1800}_{-1800}$ & $144^{+24}_{-27}$  \\
$\mu$ (mag~arcsec$^{-2}$) & $27.9$ & $26.9$ \\
\feh (dex) & $-1.8$ & $-1.5$  \\
$E(B-V)$ & 0.124 & 0.113  \\
$\ell$ (deg) & 300.265 & 14.188 \\
$b$ (deg) & 21.902 & 30.289 \\
$D_{\rm GC}$ (kpc) & 112.7 & 12.9  \\[-0.5em]
\multicolumn{3}{c}{\hrulefill} \\
$\mu_{\alpha} \cos \delta$ (mas~yr$^{-1}$) & $0.00^{+0.19}_{-0.18}$ & $-1.7^{+0.4}_{-0.4}$  \\
$\mu_{\delta}$ (mas~yr$^{-1}$) & $-0.46^{+0.25}_{-0.26}$ & $-1.6^{+0.2}_{-0.2}$ \\
$\sum_i p_{i, {\rm MM}}$ & $15.0_{-1.6}^{+1.7}$ & $4.2_{-4.2}^{+1.7}$  \\[+0.5em]
\enddata
\tablecomments{Uncertainties were derived from the highest density interval containing the peak and 68\% of the marginalized posterior distribution.}
\tablenotetext{a}{We assume a systematic uncertainty of $\pm0.1$ associated with isochrone modeling.}
\tablenotetext{b}{The age posterior peaks at the upper bound of the allowed parameter range ($13.5\Gyr$); thus, we quote a lower limit at the 84\% confidence level.}
\tablenotetext{c}{The uncertainty in $M_V$ was calculated following \citet{Martin:2008} and does not include uncertainty in the distance.}
\vspace{-3em}
\end{deluxetable}

\subsection{Morphological and Isochrone Parameters}
\label{sec:morph}

We fit the morphological and isochrone parameters of \habanero and \cayenne using the maximum likelihood formulation implemented in the ultra-faint galaxy likelihood toolkit \citep[\ugali\footnote{\url{https://github.com/DarkEnergySurvey/ugali}};][]{Bechtol:2015,Drlica-Wagner:2015,PaperI}.
The spatial distribution of stars was modeled with a \citet{Plummer:1911} profile, and a synthetic isochrone from \citet{Bressan:2012} was fit to the observed color--magnitude diagram.
We simultaneously fit the R.A. and decl. (\ra and \dec, respectively), extension ($a_h$), ellipticity (\ellip), and position angle (P.A.) of the Plummer profile, and the age (\age), metallicity (\metal), and distance modulus shift (\modulus) of the isochrone.
The posterior probability distributions of each parameter were derived using an affine-invariant Markov Chain Monte Carlo (MCMC) ensemble sampler \citep[\code{emcee}; ][]{Foreman-Mackey:2013}.
\tabref{properties} presents the best-fit parameters with uncertainties for both objects.
From these properties, we derive estimates of the Galactocentric longitude and latitude ($\ell$ and $b$, respectively), the azimuthally averaged angular and physical half-light radii ($r_h$ and $r_{1/2}$, respectively), the heliocentric distance ($D_{\odot}$), the Galactocentric distance ($D_{\rm GC}$; calculated from the three-dimensional physical separation between each object and the Galactic center, assumed to be at $R_{\rm GC} = 8.178\kpc$; \citealt{Abuter:2019}), the average surface brightness within one half-light radius ($\mu$), the stellar mass integrated along the isochrone ($M_{*}$), and the metallicity (\feh).
The \ugali membership probability ($p_{\rm \ugali}$) of each star was calculated from the Poisson probabilities to detect that star based upon its spatial position, measured flux, photometric uncertainty, and the local imaging depth, given a model that includes a putative dwarf galaxy and empirical estimation of the local stellar field population.
We define the sum of \ugali membership probabilities as $\sum_{i} p_{i, {\rm \ugali}}$.
Note that, due to incomplete coverage (i.e., the inhomogeneous background) in the region around \cayenne, it is possible that our characterization of of its parameters may be slightly biased.
However, based on the best-fit half-light radius and predicted Plummer profile (\figref{cayenne_membership}), we expect that only $3\pm2$ likely member stars lie outside our covered region (compared to $\sum_i p_{i, {\rm \ugali}} = 50_{-9}^{+8}$ for \cayenne).

\habanero was significantly detected in this likelihood analysis with a test statistic (${\rm TS}$) of ${\rm TS} = 308.4$, corresponding to a Gaussian significance of $17.6 \sigma$ (a discussion of the likelihood formalism used here is presented in Appendix~C of \citealt{PaperI}).
\cayenne was detected at ${\rm TS} = 146.7$, or $12.1 \sigma$, which is more significant than many other satellites in DES data \citep{Bechtol:2015,Drlica-Wagner:2015}.
We note that \cayenne was simultaneously discovered at a lower significance in data from Pan-STARRS PS1 \citep{PaperI}, lending confidence to the reality of this system.
While \citet{PaperI} measured a smaller half-light radius, a larger absolute magnitude, and a smaller stellar mass for \cayenne, these discrepancies are within reported uncertainties and likely explained by the difference in depth between the early DELVE-WIDE and Pan-STARRS DR1 datasets.

The spatial distributions and color--magnitude diagrams of stars in $0\fdg5 \times 0\fdg5$ regions around the centroids of \habanero and \cayenne are shown in the left two panels of \figrefs{habanero_membership}{cayenne_membership}, respectively.
The membership probabilities for individual stars are computed using the spatial and initial mass function probabilities and isochrone selection from \ugali.
Stars with $p_{\rm \ugali} > 5\%$ are colored by their membership probability, and stars with $p_{\rm \ugali} \leq 5\%$, which are almost certainly Milky Way foreground stars, are shown in gray.
The measured size and brightness suggest that \habanero is likely an ultra-faint galaxy, while \cayenne is likely a faint star cluster in the Milky Way halo.
Characteristics of each system are discussed in detail in \secref{discussion}.

\subsection{Proper Motion}
\label{sec:motion}

To see if stars in each system show coherent systemic motion on the sky, we cross-matched stars in the DELVE-WIDE catalog to the \Gaia DR2 catalog \citep{Gaia:2018} to measure their proper motions.
The stellar sample was filtered by selecting stars consistent with zero parallax ($\varpi - 3 \sigma_\varpi \leq 0$) and small proper motions (i.e., removing stars that would be unbound to the Milky Way if they were at the distance of a given system). %, and an old metal-poor isochrone.
Stars were selected within 1\fdg0 and 0\fdg5 for \habanero and \cayenne, respectively, based on a color--magnitude selection of 0.1\magn in $g$--$r$ from a best-fit isochrone with metallicity $\metal=0.0002$ and age $\age=13.5\Gyr$ for \habanero, and $\metal=0.0005$ and $\age=12.5\Gyr$ for \cayenne; this color selection was expanded to 0.2\magn for the main-sequence turnoff in \cayenne.
We note that \Gaia DR2 has a limiting magnitude of $G \sim 21\magn$ \citep{Gaia:2018}, which is significantly shallower than that of the DELVE-WIDE dataset.

For the selected stellar sample, we applied a Gaussian mixture model to determine the proper motions of the satellite candidates while accounting for the Milky Way foreground \citep{Pace:2019}.
Briefly, the mixture model separates the likelihoods of the satellite and the Milky Way stars, decomposing each into a product of spatial and proper motion likelihoods.
Stars that are closer to the centroid are given higher weight by assuming the best-fit projected Plummer profile (from \secref{morph}), and stars well outside the satellite help determine the Milky Way foreground proper motion distribution.
The MultiNest algorithm \citep{Feroz2008MNRAS.384..449F, Feroz2009MNRAS.398.1601F} was used to determine the best-fit parameters, including the proper motions of the satellite and of the Milky Way foreground stars.
The mixture model membership probability ($p_{\rm MM}$) of each star was calculated by taking the ratio of the satellite likelihood to the total likelihood from the posterior distribution (see \citealt{Pace:2019} for more details).

We derive a proper motion for \habanero of $(\mu_{\alpha} \cos \delta, \mu_{\delta}) = (0.00^{+0.19}_{-0.18}, -0.46^{+0.25}_{-0.26}) \mas~\yr^{-1}$ (\tabref{properties}, right panel of \figref{habanero_membership}) and a proper motion for \cayenne of $(\mu_{\alpha} \cos \delta, \mu_{\delta}) = (-1.7^{+0.4}_{-0.4}, -1.6^{+0.2}_{-0.2}) \mas~\yr^{-1}$ (\tabref{properties}, right panel of \figref{cayenne_membership}).
In the right panels of \figrefs{habanero_membership}{cayenne_membership}, stars with $p_{\rm MM} > 5\%$ are colored by their membership probability, and stars with $p_{\rm MM} \leq 5\%$, which are almost certainly Milky Way foreground stars, are shown in gray.
Stars cross-matched between DELVE-WIDE and \Gaia DR2 with $p_{\rm MM} > 5\%$ are outlined in the center panels of \figrefs{habanero_membership}{cayenne_membership}.
We define the sum of the mixture model membership probabilities as $\sum_{i} p_{i, {\rm MM}}$.
We find $\sum_{i} p_{i, {\rm MM}} = 15.0_{-1.6}^{+1.7}$ and $\sum_i p_{i, {\rm MM}} = 4.2_{-4.2}^{+1.7}$ for members with proper motions consistent with \habanero and \cayenne, respectively.

Based on the posterior distributions, number of stars, and diagnostic plots, we clearly detect the proper motion of \habanero, helping confirm that it is a real system.
While we do not find enough member stars to robustly disentangle the proper motion of \cayenne from the Milky Way foreground, the lack of a clear proper motion detection in \Gaia DR2 for \cayenne does not disqualify it as a real stellar system.
Importantly, \Gaia DR2 has a limiting magnitude of $G \sim 21$\magn \citep{Gaia:2018}, while the most probable member stars of \cayenne are old main-sequence stars fainter than $G \sim 21$\magn, according to the \ugali analysis.
If we assume that \cayenne has a \citet{Chabrier:2001} initial mass function with an age of $12.5\Gyr$ and \feh of $-1.5$ dex, then we predict that we should observe $N=6\pm3$ stars brighter than $G\sim21$\magn based on 1000 \ugali simulations.
Performing a similar calculation for \habanero with an age of $13.5\Gyr$ and \feh of $-1.8$ dex, we predict that we should observe $N=22\pm5$ stars.
%In both cases, the simulations predict more stars than we observed; \FIXME{we suspect that this is partially due to the relatively high amount of Milky Way foreground reddening for each of these satellites.}
Given the small number of predicted members accessible at these brighter magnitudes, it is unsurprising that there is no clear proper motion signal for \cayenne, which has far fewer likely member stars than \habanero, in \Gaia DR2.

%-------------------------------------------------------------------------------

\section{Discussion}
\label{sec:discussion}

We have presented the discovery of two new stellar systems and characterized their morphology, stellar age and metallicity, distance, and kinematics.
These measurements can provide insight into their likely natures as a dark-matter-dominated faint dwarf galaxy satellite and a faint halo star cluster, respectively.

\begin{figure*}
\center
\includegraphics[width=\textwidth]{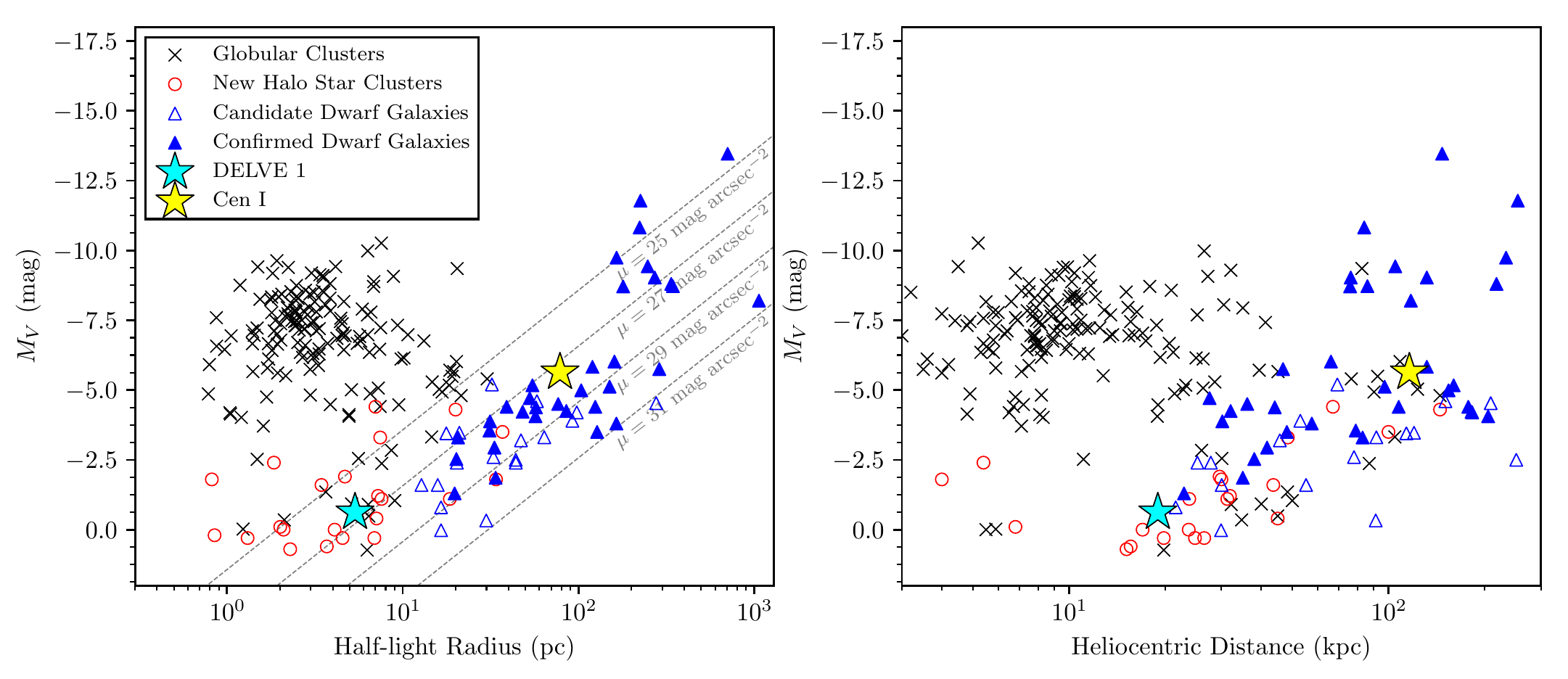}
\caption{
    (Left) Absolute magnitude vs. azimuthally averaged physical half-light radius of Milky Way dwarf galaxy satellites \citep[unfilled and filled blue triangles for candidate and confirmed dwarf galaxies, respectively;][and references therein]{PaperI}, globular clusters \citep[black crosses;][]{Harris:1996}, and recently discovered Milky Way halo star clusters \citep[unfilled red circles;][]{Fadely:2011,Munoz:2012,Balbinot:2013,Belokurov:2014,Laevens:2014,Kim:2015a,Laevens:2015b,Kim:2016,Luque:2016,Luque:2017,Luque:2018,Koposov:2017,Mau:2019,Torrealba:2019}.
    \habanero is shown as a yellow star, and \cayenne is shown as a cyan star.
    Lines of constant surface brightness are drawn as dashed gray lines.
    (Right) Absolute magnitude vs. heliocentric distance of stellar systems in orbit around the Milky Way.
    \habanero occupies a position in this three-dimensional parameter space consistent with the population of ultra-faint galaxy satellites of the Milky Way, while the small physical size and heliocentric distance of \cayenne are more consistent with those of faint halo star clusters.
}
\label{fig:satellites}
\end{figure*}

The left panel of \figref{satellites} presents the distribution of Milky Way dwarf galaxy satellites (unfilled and filled blue triangles), Milky Way halo star clusters (unfilled red circles), and globular clusters (black crosses) in size--luminosity space, and the right panel of \figref{satellites} shows the same satellites in distance--luminosity space.
Based on their positions in \figref{satellites}, \habanero appears to have properties that are consistent with the population of known dwarf galaxy satellites of the Milky Way (e.g., its properties are similar to those of Leo~IV), and \cayenne appears to have properties that are consistent with the population of known halo star clusters of the Milky Way (e.g., its properties are similar to those of Mu\~noz~1).
However, further investigations will be needed to confirm these classifications.
The derived properties of each object are discussed in detail in the following subsections.

\subsection{Centaurus~I}

We found that \habanero is an old ($\age>12.85\Gyr$), extended ($r_{1/2}=79^{+14}_{-10}\pc$), and faint ($M_V=-5.55^{+0.11}_{-0.11}$ \magn) stellar system with an average systemic metallicity ($\feh = -1.8$ dex) consistent with that of most ultra-faint galaxies \citep[][2019 edition]{McConnachie:2012}.
Given its physical size, \habanero is relatively bright compared to the population of ultra-faint galaxies with similar size, but its absolute magnitude and physical size are consistent with the definition for ultra-faint galaxies put forth in \citet{Simon:2019}; i.e., a dwarf galaxy with $M_V \gtrsim -7.7\magn$.

The well-populated horizontal branch of \habanero makes it an excellent candidate for RR Lyrae star searches.
In particular, Equation~4 of \citet{Martinez-Vazquez:2019} predicts that a system with $M_V=-5.55\magn$ should have $\gtrsim 6$ RR Lyrae stars.
%SM: The calculation gives 6.45
%In particular, other ultra-faint dwarfs with absolute magnitudes similar to that of \habanero have $\roughly 5$--12 RR Lyrae stars (see Table~A1 and Figure~10 in \citealt{Martinez-Vazquez:2019} for details).
Discovering RR Lyrae stars in \habanero would aid in verifying the nature of this system by allowing for the determination of its physical properties with greater precision \citep[e.g.,][]{Greco:2008,Garofalo:2013,Vivas:2016,Ferguson:2019}.

Investigation of the region around \habanero reveals a secondary, less significant overdensity displaced  $\roughly 0\fdg17$ to the west of \habanero (red circle in \figref{habanero_diagnostic}).
This elongated overdensity near the centroid of \habanero could be a candidate tidal feature of the system.
Such potentially tidally disrupted structures have previously been observed in some Milky Way satellites \citep[e.g.,][]{Sand:2009,Munoz:2010,Sand:2012,Roderick:2015} and provide clues to investigating the dynamical state of these systems \citep[e.g.,][]{Piatek:1995,Deason:2012,Lokas:2012,Collins:2017}.
For instance, the fact that the tidal tails in Tucana~III show a high velocity gradient, but no significant density variation, suggests that it is on radial orbit and had a recent close pericentric passage about the Milky Way \citep{Drlica-Wagner:2015,Li:2018}.
This radial orbit is confirmed by dynamical modeling \citep{Erkal:2018} and proper motion measurements from \Gaia DR2 \citep{Simon:2018}.
%Examples of candidate tidal features are unusually high ellipticites \citep[e.g., Hercules, Ursa~Major~II;][]{Sand:2009,Munoz:2010}, irregular outer isophotes \citep[e.g., Ursa~Major~I, Ursa~Major~II;][]{Okamoto:2008,Munoz:2010}, extra-tidal sub-structures \citep[e.g., Hercules][]{Sand:2009}, and kinematic sub-structure or velocity gradients \citep[e.g., Coma~Berenices, Hercules, Leo~V, Ursa~Major~II;][]{Simon:2007,Aden:2009,Collins:2017}.
%It is important to note that simulations of tidal stripping do not generally reproduce these features \citep{Munoz:2008}.
%\YC{Transition between the above two sentences is abrupt.}

However, this might not be the case for \habanero.
Even if a very eccentric orbit is assumed for \habanero, its current location is too far from the center of the Milky Way to maintain features induced by tidal stripping after a close pericenter \citep[e.g.,][]{Penarrubia:2008,Kazantzidis:2011,Barber:2015}.
Tidal structures induced during the pericentric passage seem to be short-lived and are expected to fade out while traveling to the apocenter.
According to \citet{Li:2018}, circumstantial morphological properties alone cannot provide reliable evidence for tidal features.
We also note that the displacement of the secondary overdensity is not aligned with the solar-reflex-corrected proper motion of \habanero; this is reminiscent of the Hercules ultra-faint galaxy, which exhibits elongated and irregular morphology perpendicular to a very eccentric orbit at a heliocentric distance of 140\kpc \citep{Kupper:2017}.
\citet{Garling2018ApJ...852...44G} used observations of RR Lyrae variable stars to determine that much of the stellar content of Hercules has been stripped, with its orbit aligned along its minor axis.
Other recent studies have been inconclusive about whether or not Hercules has undergone tidal stripping \citep{Fu:2019}.
In addition, follow-up deep imaging has shown that candidate tidal features identified in relatively shallow imaging surveys can actually be artifacts caused by clumps of Milky Way foreground and background stars \citep[e.g., Leo~V;][]{Mutlu-Pakdil:2019}.
Given that the secondary overdensity appears to be disconnected from the centroid of \habanero (\figref{habanero_diagnostic}), it is also conceivable that it is an associated companion \citep[e.g., as with Car~II and Car~III;][]{Torrealba:2018a}.
Thus, follow-up deep imaging and spectroscopic studies of the candidate tidal features of \habanero are needed to illuminate their structure and determine whether this secondary overdensity is indeed physically associated with \habanero.

\begin{figure*}
\center
\includegraphics[width=\columnwidth]{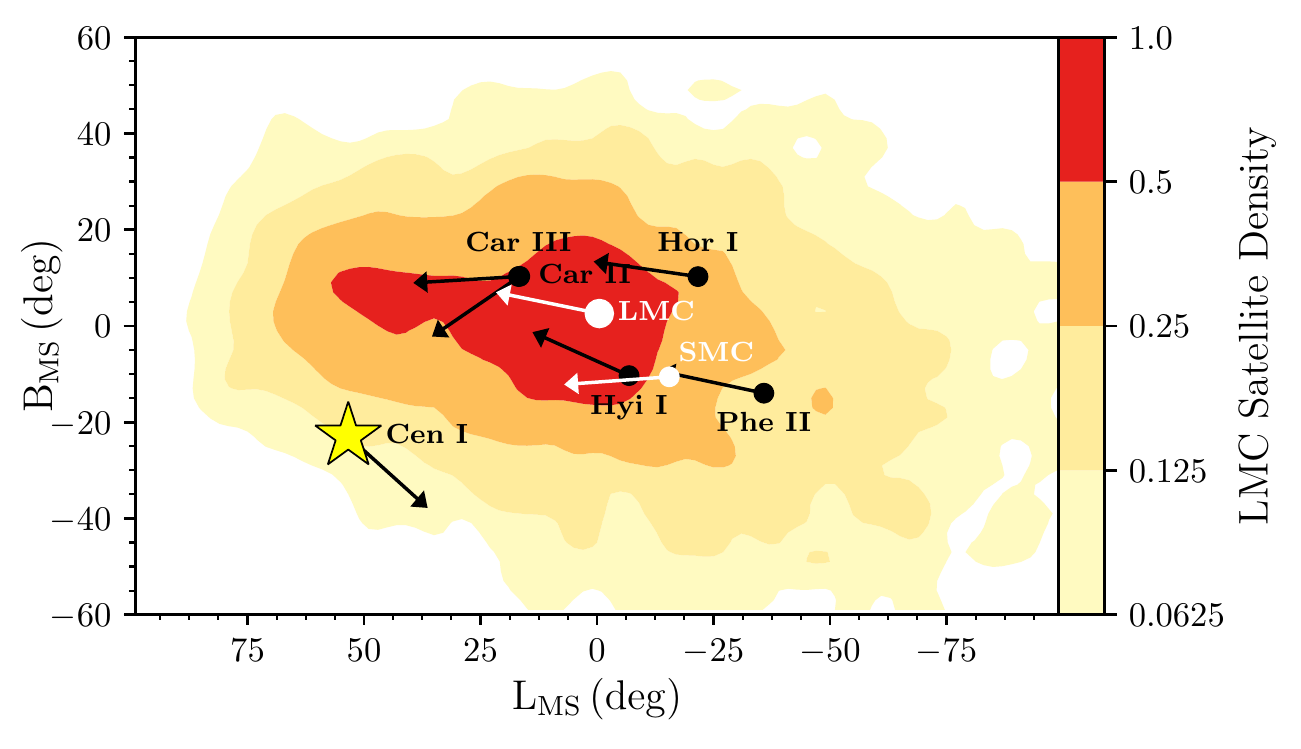}
\caption{
    Relative density of simulated LMC satellites from \citet{Jethwa:2016} normalized to unity.
    \habanero is shown as a yellow star, and five likely LMC satellites \citep[Hor~I, Car~II, Car~III, Hyi~I, Phe~II;][]{Kallivayalil:2018,Erkal:2019b} are shown as black circles; the LMC and SMC are shown as white circles.
    Arrows indicate the solar-reflex-corrected proper motions of each system (no physical meaning is attributed to the magnitudes of these arrows).
    Note that Car~II and Car~III are spatially coincident but have different proper motion vectors.
    The motion and position of \habanero are opposite to those of the LMC and its satellites, making an association unlikely.
    \cayenne does not appear because it is located at ($L_{\rm MS}$, $B_{\rm MS}$) = $(135\deg, -67\deg)$.
}
\label{fig:lmc}
\end{figure*}

It is also interesting to consider the possible origins of \habanero.
DES has revealed a concentration of Milky Way ultra-faint galaxy satellites around the Large and Small Magellanic Clouds (the LMC and SMC, respectively), suggesting that the LMC has brought its own satellite population into the Milky Way \citep[e.g.,][]{D'Onghia:2008a,Deason2015MNRAS.453.3568D,Sales2017MNRAS.465.1879S,Jethwa:2018, Kallivayalil:2018,Erkal:2019b,Jahn:2019,PaperII}.
Hence, we consider whether or not \habanero is associated with the LMC, given their relatively small angular separation ($\roughly 58\degree$).
To investigate the potential association of \habanero with the LMC, we present the spatial position and solar-reflex-corrected proper motion vector of \habanero over simulated LMC tidal debris from \citet{Jethwa:2016} in Magellanic Stream coordinates \citep{Nidever:2008} along with the LMC, SMC, and five other ultra-faint galaxies associated with the LMC in \figref{lmc}.
These five ultra-faint galaxies are Horologium~I, Carina~II, Carina~III, Hydrus~I, and Phoenix~II \citep{Kallivayalil:2018,Erkal:2019b}, with proper motion measurements coming from \citet{Kallivayalil:2018} and \citet{Pace:2019}.
While \citet{Pardy:2019} suggested that Carina and Fornax are satellites of the LMC, orbit modeling done by \citet{Erkal:2019b} found that neither Carina nor Fornax are likely LMC satellites; hence, we do not include these systems in \figref{lmc}.
The position of \habanero is only marginally consistent with that of the simulated LMC satellites, and its proper motion is nearly antiparallel to that of the LMC and its probable satellites.
Thus, we find no strong evidence in support of \habanero being a satellite of the LMC from this analysis.

\citet{Erkal:2019b} put forward an alternative technique for determining LMC membership where the orbit of each satellite is rewound in the presence of the Milky Way and LMC to determine if they were bound to the LMC before it fell onto the Milky Way.
\citet{Erkal:2019b} also used this technique on satellites without radial velocity measurements to determine if there were any radial velocities for which the satellites belonged to the LMC.
This is done by sampling the proper motions and distances from their observed uncertainties while sampling the radial velocity uniformly from $-500$ to $500\kms$.
This sampling was done 100,000 times, and, for each realization, we rewound \habanero in the combined presence of the LMC and the Milky Way for 5\Gyr to determine whether it was originally bound to the LMC.
In this analysis, we model the LMC as a Hernquist profile \citep{Hernquist:1990} with a mass of $1.5\times10^{11} M_{\odot}$ and a scale radius of $17.13\kpc$, consistent with recent measurements of the LMC mass \citep{Penarrubia:2016,Erkal:2019a}.
With this analysis, we find that, for radial velocities between $350$ to $410\kms$, \habanero has a $>5\%$ chance of being an LMC satellite with a maximum probability of $\roughly 10\%$ at a radial velocity of $385\kms$.
While this probability is still small, if the radial velocity is found to be in this range, it would warrant additional investigation.
Outside of this range, the probability quickly drops below $1\%$ for radial velocities below $\roughly 300\kms$ and above $490\kms$.
We also note that future proper motion measurements with \Gaia DR3 will improve the proper motion uncertainties and thus give a more accurate trajectory when rewinding the orbit of \habanero.

We also consider whether \habanero is associated with the Vast Polar Structure (VPOS) of the Milky Way \citep{Pawlowski:2012}.
A large fraction of Milky Way satellite galaxies have recently been determined \citep{Fritz:2018} to lie on a thin, corotating plane nearly perpendicular to the Milky Way's stellar disk \citep{Pawlowski:2013,Pawlowski:2019}.
Adopting the same VPOS parameters as \citet{Fritz:2018}, namely the assumed normal $(l_{\rm MW},b_{\rm MW})=(169.3,-2.8)\deg$ and angular tolerance $\theta_{\rm inVPOS} = 36\fdg87$, we find it unlikely that \habanero is a VPOS member.
Specifically, the minimum possible angle between the VPOS and the satellite's orbital pole based on spatial information alone is $\theta_{\rm pred} = 35\fdg15$, while the probability that the orbital pole lies within $\theta_\text{inVPOS}$ of the VPOS normal ranges from $\roughly 4$--$10\%$ depending on the assumed heliocentric radial velocity.
In addition, the available spatial and proper motion measurements prefer a counter-orbiting orientation relative to the VPOS.
However, we note that the orbital inconsistency does not rule out the possibility of \habanero being a VPOS member as this analysis is based on the limited information currently available.
A radial velocity measurement is required to conclusively categorize \habanero as either a VPOS member or not and to determine whether or not it is co- or counter-orbiting.

\subsection{DELVE~1}

We identified \cayenne as a faint ($M_V=-0.2^{+0.8}_{-0.6}$ \magn), compact ($r_{1/2}=5.4^{+1.5}_{-1.1}\pc$), and low-mass (${\rm M}_{\star}=144^{+24}_{-27}\Msun$) stellar system located at a relatively close heliocentric distance (${\rm D}_{\odot}=19.0^{+0.5}_{-0.6}\kpc$).
As such, it appears to be consistent with the population of faint halo star clusters of the Milky Way discovered in recent years \citep[e.g.,][]{Fadely:2011,Munoz:2012,Balbinot:2013,Belokurov:2014,Laevens:2014,Kim:2015a,Laevens:2015b,Kim:2016,Luque:2016,Luque:2017,Koposov:2017,Luque:2018,Mau:2019,Torrealba:2019}.

%While the population of dwarf galaxy satellites of the Milky Way has been studied in detail \citep{Simon:2019,PaperI,PaperII}, the growing population of faint halo star clusters warrants further investigation \citep[however, see][]{Boylan-Kolchin;2017,Forbes:2018,El-Badry:2018}.
%The observed faint halo star clusters are currently characterized by their small physical sizes ($r_{1/2} \lesssim 20\pc$), relatively small heliocentric distances (${\rm D}_{\odot} \gtrsim 15\kpc$), and faint surface brightnesses ($\mu \gtrsim 25 \magn \asec^{-2}$).
%They are fainter than most classical globular clusters while being more compact than most ultra-faint galaxies at similar surface brightness (\figref{satellites}).
%They also exhibit no convincing evidence for significant dark matter content and/or extended star formation histories \citep{Simon:2007,Tollerud:2010,Willman:2012}.
%%{\bf (TSL:I think another reference is needed to show that these star clusters have lower DM content or low FeH dispersion. I do not think we know that. I agree the globular clusters are.)} %ADW: +1

These faint halo clusters have been proposed to be the remnants of merger events---they were accreted onto the Milky Way along with their host galaxies, but the host galaxies themselves were disrupted due to the Milky Way tides \citep[e.g.,][]{Searle:1978,Gnedin:1997, Koposov:2007,Forbes:2010,Leaman:2013,Massari:2017}.
Specifically, the compactness of these star clusters is essential to longer survival timescales despite the strong tidal fields during merging processes, while the host galaxies of these clusters are disrupted by the Milky Way tides on shorter timescales.
This scenario has received considerable observational support from the age, metallicity, and spatial distributions of these clusters \citep[e.g.,][]{Zinn:1993a,DaCosta:1995, Mackey:2004,Marin-Franch:2009,Dotter:2010,Mackey:2010,Keller:2011} and is further supported by the close resemblance between Milky Way halo clusters and the clusters thought to have been accreted with the dwarf galaxies that fell into the Milky Way \citep[e.g.,][]{Smith:1998,Johnson:1999,DaCosta:2003,Wetzel:2015,Yozin:2015,Bianchini:2017}.
Meanwhile, with the advent of \Gaia, the assembly history of the Milky Way has been revealed in greater detail, shedding light on the origin of these systems.
Recently, kinematic data from \Gaia have been used to propose that $\roughly 35\%$ of the Milky Way globular clusters were accreted with merger events \citep{Massari:2019}.
In addition, \citet{Kruijssen:2019} suggested that $\roughly40\%$ of the Milky Way globular clusters formed ex situ and accreted through merger events based on analysis of the age--metallicity distributions of the globular clusters.
This proposal has been supported by a chemical abundance analysis of Palomar~13, which found possible similarities between Palomar~13 and other globular clusters that purportedly accreted through either the \Gaia-Enceladus or Sequoia events \citep{Koch:2019}.

Although \cayenne does not have spectroscopically measured metallicity or radial velocity, it is possible to consider whether \cayenne was accreted with the LMC, which is known to have brought a large population of star clusters \citep{Bica:2008}.
The age and metallicity of \cayenne are consistent with those found in the LMC star clusters; however, the position of \cayenne in the sky easily rules out its association with the LMC, even if the leading arm of the MC (a very extended H I gas structure) is taken into account \citep{Nidever:2010}.
%We also find that \cayenne is very unlikely to be a VPOS member unless it has a high ($>500\km/\second$) radial velocity.
It is important to note that, without spectroscopic information, we are unable to draw a robust conclusion on the origin of \cayenne.

Exploring the long-term dynamical evolution of \cayenne may also provide insights into its origins.
To estimate its survival timescale in its current evolutionary state, we compute the evaporation timescale (i.e., the time over which stars in a star cluster escape the system due to two-body relaxation) following \citet{Koposov:2007}.
Specifically, we compute $t_{\rm ev} \simeq 12 t_{\rm rh}$ \citep{Koposov:2007}, where $t_{\rm rh}$ is the half-mass relaxation time given by Equation~7.2 of \citet{Meylan:1997}:
\begin{equation*}
t_{\rm rh} = 0.138 \frac{M^{1/2} R_h^{3/2}}{\langle m \rangle G^{1/2} \ln\Lambda},
\label{eqn:tev}
\end{equation*}
where $M$ is the total stellar mass of the cluster, $R_h$ is the half-mass radius (we assume $R_h \sim r_{1/2}$), $\langle m \rangle$ is the mean stellar mass of stars in the cluster, $G$ is the gravitational constant, and $\Lambda \simeq 0.4 N$, where $N$ is the total number of stars in the cluster (a richness of 610 was fit to \cayenne with \ugali).
We find $t_{\rm ev} = 3\Gyr$ for \cayenne, which is longer than that of both Kop~1 and Kop~2 (0.7\Gyr and 1.1\Gyr, respectively; \citealt{Koposov:2007}).
However, this evaporation timescale is still only a quarter of the estimated lifetime of the star cluster ($\age = 12.5\Gyr$), suggesting that \cayenne cannot have persisted in its observed structural and dynamic state throughout its lifetime.

\subsection{Classification of Ultra-faint Objects}

As noted above, the physical classifications of \habanero and \cayenne are uncertain without spectroscopic information.
%Such a situation has become common as large digital sky surveys have extend to fainter magnitudes and surface brightnesses.
Classifications based on physical size and absolute magnitude have become less certain as surveys have revealed a continuum of objects located between the size--luminosity loci of classical dwarf galaxies and globular clusters. 
In particular, the classification of systems with $M_V > -2 \magn$ and $10 \pc \lesssim r_{1/2} \lesssim 40\pc$ is uncertain, leading authors to call this region of parameter space the ``valley of ambiguity'' \citep{Gilmore:2007,Conn:2018a,Conn:2018b}.

While both \habanero and \cayenne reside outside the most ambiguous region of parameter space, definitive classification rests on the determination of the dynamical mass by measuring the velocity dispersion.
Generally, for an ultra-faint satellite, a resolved velocity dispersion implies the presence of a dark matter halo and, definitionally, a classification as a dwarf galaxy \citep{Willman:2012}.
However, despite significant investment of telescope time, observations of many recently discovered systems lack sufficient statistical and systematic precision to resolve a velocity dispersion smaller than a few $\kms$ \citep[e.g.,][]{Simon:2019}.
Many newly discovered Milky Way satellites still lack clear velocity and/or metallicity dispersion measurements \citep{Kirby:2015a,Kirby:2017,Martin:2016a,Martin:2016b,Walker:2016,Simon2017ApJ...838...11S}, making it difficult to reliably categorize them as either faint star clusters or ultra-faint dwarfs.
%The lack of a resolved velocity dispersion may indicate that these systems are star clusters ($\roughly 0.1$--$0.2\kms$); however, it may also be due to the low number of stars, a velocity dispersion that is smaller than the systematic errors, or that the .
This has led to the adoption of other indirect arguments to infer the presence of a dark matter halo, including large metallicity dispersions \citep{Simon2011,Willman:2012}, lack of light element correlations \citep[e.g., in Tucana~III;][]{Marshall2019ApJ...882..177M}, and/or low neutron-capture element abundances \citep{Ji2019ApJ...870...83J}.
%, large size \citep[e.g., Tuc~III;][]{Simon2017ApJ...838...11S},
These indirect classification criteria are founded on the argument that only systems with a dark matter halo are able to retain and self-enrich their gas after the initial episodes of star formation \citep[e.g.,][]{Kirby:2013} and generally rely on the lack of star clusters with these observed properties.
% One issue with these arguments is that the classification scheme has only been tested on the bright, well-studied objects (i.e., pre-SDSS star clusters and the bright dwarf galaxies). 
% The faint outer halo star clusters generally lack spectroscopic follow-up.
However, even metallicity arguments are challenging when there are few member stars that are bright enough for spectroscopic follow-up with current facilities.

%However, unambiguously distinguishing these two classes of ultra-faint objects is challenging because their extremely low luminosities make it difficult to obtain metallicity and radial velocity measurements of member stars.
%More significant metallicity spreads and radial velocity dispersions are expected in galaxies compared to star clusters.
%However, even these metallicity arguments are challenging when there are few member stars that are bright enough for spectroscopic follow-up with current facilities, and current instruments cannot fully resolve velocity dispersion for these ultra-faint objects.
The classification challenge will become more pressing in the coming decade with the advent of the Large Synoptic Survey Telescope (LSST), which is expected to discover up to several hundred new ultra-faint galaxy candidates and an as-of-yet unpredicted number of faint halo star clusters \citep{Drlica-Wagner:2019}.
Upcoming 30 m telescopes will provide access to the spectra of fainter member stars, but instrument stability will likely still be a driving limitation in resolving the small velocity dispersions expected in these systems.
Understanding how to classify new ultra-faint stellar systems will thus be an important and challenging issue in the era of LSST, particularly when using the population demographics of the Milky Way ultra-faint galaxies as a probe of dark matter microphysics.
In the end, even with all available information, it still may only be possible to make probabilistic classifications of these systems, which can be folded into studies of the Milky Way’s ultra-faint galaxy population as a systematic uncertainty.

\section{Summary}
\label{sec:conclusion}

We present the discovery of two ultra-faint stellar systems, \habanero and \cayenne, in early data from the DELVE survey.
These stellar systems were the most significant new stellar overdensities detected in an automated search of $\roughly 6{,}000\deg^2$ in the southern hemisphere.
%DELVE is an ongoing DECam program with 126 allocated nights, observing from 2019A to 2021B.
%It has three components: WIDE survey, DEEP survey, and the Magellanic Cloud survey.
%As a part of the DELVE-WIDE program, we constructed a catalog-level coadd using the early DELVE exposures (mostly from 2019A observing runs) and the existing DECam exposures to search for new Milky Way satellites.
%From the analysis of stellar overdensities in these early DELVE-WIDE data, which cover $\roughly 6000\deg^2$ area in the southern hemisphere, we discovered two new ultra-faint stellar systems.
Based on morphological and isochrone modeling, we tentatively classify \habanero as an ultra-faint galaxy and \cayenne as a faint halo star cluster.
Using proper motions from \Gaia DR2, we confirmed that both of these systems appear to be physically bound associations of stars with coherent motion on the sky.
We also found that neither of these satellites is likely to be associated with the LMC and that \habanero is unlikely to be associated with the VPOS.
Given these two discoveries in the early DELVE-WIDE data and predictions from numerical simulations \citep[e.g.,][]{Nadler:2018}, we anticipate that DELVE will discover $\roughly 10$ satellite galaxies as it continues to complete contiguous DECam coverage of the southern sky.
Furthermore, \citet{PaperII} predicted that $\roughly 100$ satellites of the Milky Way with $M_V < 0\magn$ and $r_{1/2} > 10\pc$ still remain to be discovered, and DECam surveys like DELVE will play an important role in advancing this census.

%-------------------------------------------------------------------------------

\section{Acknowledgments}

S.M. is supported by the University of Chicago Provost's Scholar Award.
A.B.P. acknowledges support from NSF grant AST-1813881.
J.L.C. acknowledges support from NSF grant AST-1816196.
Research by D.J.S. is supported by NSF grants AST-1821987, AST-1821967, AST-1813708, AST-1813466, and AST-1908972.
A.M. acknowledges support from CONICYT FONDECYT Regular grant 1181797.
This project is partially supported by the NASA Fermi Guest Investigator Program Cycle 9 No.\ 91201.
This work is partially supported by Fermilab LDRD project L2019-011.

% DECam acknowledgement approved by DES and NOAO.
% LaTeX 2e version generated 2017-Dec-27 
% shaw@noao.edu

This project used data obtained with the Dark Energy Camera (DECam), 
which was constructed by the Dark Energy Survey (DES) collaboration.
Funding for the DES Projects has been provided by 
the DOE and NSF (USA),   
MISE (Spain),   
STFC (UK), 
HEFCE (UK), 
NCSA (UIUC), 
KICP (U. Chicago), 
CCAPP (Ohio State), 
MIFPA (Texas A\&M University),  
CNPQ, 
FAPERJ, 
FINEP (Brazil), 
MINECO (Spain), 
DFG (Germany), 
and the collaborating institutions in the Dark Energy Survey, which are
Argonne Lab, 
UC Santa Cruz, 
University of Cambridge, 
CIEMAT-Madrid, 
University of Chicago, 
University College London, 
DES-Brazil Consortium, 
University of Edinburgh, 
ETH Z{\"u}rich, 
Fermilab, 
University of Illinois, 
ICE (IEEC-CSIC), 
IFAE Barcelona, 
Lawrence Berkeley Lab, 
LMU M{\"u}nchen, and the associated Excellence Cluster Universe, 
University of Michigan, 
NSF's National Optical-Infrared Astronomy Research Laboratory, 
University of Nottingham, 
Ohio State University, 
OzDES Membership Consortium
University of Pennsylvania, 
University of Portsmouth, 
SLAC National Lab, 
Stanford University, 
University of Sussex, 
and Texas A\&M University.

This work has made use of data from the European Space Agency (ESA) mission {\it Gaia} (\url{https://www.cosmos.esa.int/gaia}), processed by the {\it Gaia} Data Processing and Analysis Consortium (DPAC, \url{https://www.cosmos.esa.int/web/gaia/dpac/consortium}).
Funding for the DPAC has been provided by national institutions, in particular the institutions participating in the {\it Gaia} Multilateral Agreement.

Based on observations at Cerro Tololo Inter-American Observatory, NSF's National Optical-Infrared Astronomy Research Laboratory (2019A-0305; PI: Drlica-Wagner), which is operated by the Association of Universities for Research in Astronomy (AURA) under a cooperative agreement with the National Science Foundation.

This manuscript has been authored by Fermi Research Alliance, LLC, under contract No.\ DE-AC02-07CH11359 with the US Department of Energy, Office of Science, Office of High Energy Physics. The United States Government retains and the publisher, by accepting the article for publication, acknowledges that the United States Government retains a non-exclusive, paid-up, irrevocable, worldwide license to publish or reproduce the published form of this manuscript, or allow others to do so, for United States Government purposes.

\facility{Blanco, \Gaia.}
\software{\code{astropy} \citep{astropy:2013,astropy:2018}, \emcee \citep{Foreman-Mackey:2013}, \code{fitsio},\footnote{\url{https://github.com/esheldon/fitsio}} \healpix \citep{Gorski:2005},\footnote{\url{http://healpix.sourceforge.net}} \code{healpy},\footnote{\url{https://github.com/healpy/healpy}} \code{Matplotlib} \citep{Hunter:2007}, \code{numpy} \citep{numpy:2011}, \code{scipy} \citep{scipy:2001}, \ugali \citep{Bechtol:2015}.\footnote{\url{https://github.com/DarkEnergySurvey/ugali}}}

\bibliography{main}

\end{document}